\documentclass[10pt]{iopart}
\usepackage{iopams}
\usepackage{graphicx}
\usepackage{bm}
\usepackage{mathrsfs}
\usepackage{iopams,setstack}
\usepackage{epigraph}
\def\qpun{\dot{q}}

\newcommand{\beq}{\begin{equation}}
\newcommand{\eeq}{\end{equation}}
\newcommand{\beqa}{\begin{eqnarray}}
\newcommand{\eeqa}{\end{eqnarray}}
\def\nn{\nonumber\\}
\def\eq#1{(\ref{#1})}

\def\qpun{\dot{q}}
\def\dfrac#1#2{\displaystyle{\frac {#1}{ #2}}}
\def\text#1{\rm #1}

\begin{document}

\title{On the global version of Euler-Lagrange equations.
}

\author{R.\ E.  Gamboa Sarav\'{\i}\dag\  and J.\ E. {Solomin}\ddag
}

\address{\dag\ Departamento de F\'{\i}sica, Facultad de Ciencias
Exactas, Universidad Nacional de La Plata, Argentina. }

\address{\ddag\ Departamento de Matem\'atica, Facultad de Ciencias
Exactas, Universidad Nacional de La Plata, Argentina. }

\ead{quique@fisica.unlp.edu.ar}



\date{\today}

\begin{abstract}
The introduction of a covariant derivative on the velocity phase
space is needed for a global expression of Euler-Lagrange
equations. The aim of this paper is to show how its torsion tensor
turns out to be involved in such a version.
\end{abstract}

\submitto{\JPA} 

\epigraph{\em The introduction of numbers as coordinates\ldots is
an act of violence} { Hermann Weyl }

An increasing attention has been recently paid to coordinate free
formulations of motion equations in Classical Mechanics (see for
instance \cite{cmr1,cmr2} and references therein). In this work we
write down intrinsic Euler-Lagrange equations and show the
appearance of a torsion term. Furthermore, we shall see that this
term should also be present in the horizontal Lagrange-Poincar\'e
equations considered in \cite{cmr1,cmr2}, if the torsion of the
chosen derivative does not vanish.

It is worth noticing that covariant derivatives with non vanishing
torsion naturally arise  in several branches of Physics; namely,
dynamics with nonholonomic constraints \cite{vf,v},  E. Cartan's
Theory of Gravity (see for instance \cite{h}) and modern string
theories (see for example \cite{p}), among others.

Let us consider a physical system with configuration manifold $Q$
and Lagrangian $L(q,\qpun): TQ\rightarrow \mathbb{R}$ (for this
geometrical setting see for instance \cite{am}).

If a coordinate free characterization  of the Euler-Lagrange
Equations associated to the system is required a covariant
derivative $D$ must be introduced on $TQ$, for $\dfrac{\partial
L}{\partial q}$ is involved (see for instance \cite{kn}). Once
such $D$ is chosen, $\dfrac{DL}{Dq}$ is defined in the standard
way \beq\label{def} \dfrac{D L}{D q}(q_0,\qpun_0)=\left
.\frac{\partial }{\partial \lambda}\right|_{\lambda=0} L
\circ\gamma(\lambda)\ ,\eeq with
$\gamma(\lambda)=(q(\lambda),\qpun_{0\parallel}(\lambda))$,
$q(0)=q_0$, $\qpun(0)=\qpun_0$ and $\qpun_{0\parallel}(\lambda)$
the parallel transport of $\qpun_0$ along $q(\lambda)$.

Moreover, an associated covariant derivative on $T^*Q$, that we
will also  denote by $D$, is naturally defined through Leibnitz
rule: for any curves $\alpha(t)$ and $v(t)$ in $T^*Q$ and $TQ$
respectively \beq \label{Leibnitz}\frac{d}{dt}\langle
\alpha(t),v(t)\rangle = \langle
\frac{D\alpha(t)}{Dt},v(t)\rangle+\langle
\alpha(t),\frac{Dv(t)}{Dt}\rangle\ ,\eeq where $\langle\ ,\
\rangle$ denotes the pairing between $T^*Q$ and $TQ$.

It is worth noticing that $\dfrac{\partial}{\partial \qpun}$ has a
coordinate free sense: it is the derivative along the fibre.

\bigskip

{\bf PROPOSITION 1.} {\em Let $D$ an arbitrary covariant
derivative on $TQ$. Then the coordinate free expression of the
Euler-Lagrange equations  is \beqa
\frac{D}{Dt}\left(\frac{\partial L}{\partial \qpun}\right) -
\frac{DL}{Dq}= \frac{\partial L}{\partial \qpun}\ T(\qpun (t),\ )\
,\eeqa where $T(\ ,\ )$  is the torsion tensor of $D$.}

{\bf Proof. } The curve $q(t)$ is a solution of the Euler-Lagrange
equations if and only if it is a critical point for the action
\beq\label{action} S= \int_{t_0}^{t_1}L(q(t),\qpun(t))dt\ ,\eeq
for variations of the curves such that $q_0$ and $q_1$ remain
fixed. That is, for each
$q(t,\lambda):[t_0,t_1]\times(-\varepsilon,\varepsilon)\rightarrow
Q$ such that $q(t,0)=q(t)$, $q(t_0,\lambda)=q(t_0)$ and
$q(t_1,\lambda)=q(t_1)$, \beq \left .\frac{\partial }{\partial
\lambda}\right|_{\lambda=0}
\int_{t_0}^{t_1}L(q(t,\lambda),\qpun(t,\lambda))dt =
\int_{t_0}^{t_1}\delta L(q(t),\qpun(t))dt=0\ ,\eeq where $\delta
L(q(t),\qpun(t))= \left .\dfrac{\partial }{\partial
\lambda}\right|_{\lambda=0}L(q(t,\lambda),\qpun(t,\lambda))$.

But \beqa \delta
L(q(t),\qpun(t))=\lim_{\lambda\rightarrow0}\frac{L(q(t,\lambda),\dot{q}(t,\lambda))-
L(q(t,0),\dot{q}(t,0))}{\lambda}\nn = \lim_{\lambda\rightarrow0}
\frac{L(q(t,\lambda),\dot{q}(t,\lambda))-
L(q(t,\lambda),\dot{q}_{\parallel}(t,\lambda))}{\lambda}\nn
+\lim_{\lambda\rightarrow0}\frac{L(q(t,\lambda),\dot{q}_{\parallel}(t,\lambda))-
L(q(t,0),\dot{q}(t,0))}{\lambda}\ ,\eeqa where
$\dot{q}_{\parallel}(t,\lambda)$ is the parallel translated of the
vector $\qpun (t,0)$ along the curve $q(t,\lambda)|_{\text{fixed}\
t}$, see Figure 1.

 Then \beqa\delta
L(q(t),\qpun(t))= \frac{\partial L}{\partial \qpun}\ D_{\delta
q(t)}\qpun(t)+ \frac{DL}{Dq}\ \delta q(t)\ ,\eeqa where we have
denoted \beqa \delta q(t)= \left .\frac{\partial
q(t,\lambda)}{\partial \lambda}\right|_{\lambda=0}\ \ \text{and}\
\  \qpun (t) = \left .\frac{\partial q(t,\lambda)}{\partial
t}\right|_{\lambda=0}\ .\eeqa

\begin{figure}\label{figura}\begin{center}
\includegraphics[height=4cm]{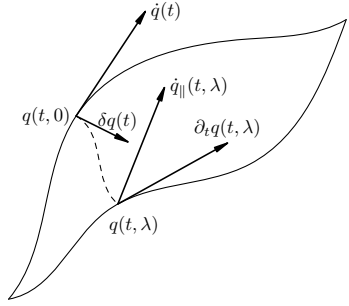}
\caption{The vector $\dot{q}_{\parallel}(t,\lambda)$ is the
parallel  transport of the vector $\qpun(t,0)$ along the dashed
curve. }\end{center}\end{figure}

By definition of the torsion tensor $T(\ ,\ )$, we have \beqa
T(\qpun (t),\delta q(t))= D_{\qpun(t)}\delta q(t) - D_{\delta
q(t)}\ \qpun(t) - [\qpun (t),\delta q(t) ]\ .\eeqa

Thus, by using \eq{Leibnitz} and taking into account that $[\qpun
(t),\delta q(t)]$ vanishes, we have \beqa \delta L(q(t),\qpun(t))=
\frac{\partial L}{\partial \qpun}\ D_{\qpun(t)}\delta q(t)+
\frac{DL}{Dq}\ \delta q(t)- \frac{\partial L}{\partial \qpun}\
T(\qpun (t),\delta q(t))\nn =\frac{d}{dt}\left(\frac{\partial
L}{\partial \qpun}\ \delta
q(t)\right)-\frac{D}{Dt}\left(\frac{\partial L}{\partial
\qpun}\right) \delta q(t) + \frac{DL}{Dq}\ \delta q(t)-
\frac{\partial L}{\partial \qpun}\ T(\qpun (t),\delta q(t))\ .\nn
\eeqa Now integrating along the curve $q(t)$  we finally get \beqa
\frac{D}{Dt}\left(\frac{\partial L}{\partial \qpun}\right) \delta
q(t) - \frac{DL}{Dq}\ \delta q(t)= \frac{\partial L}{\partial
\qpun}\ T(\qpun (t),\delta q(t))\ . \ \ \ \Box \eeqa

{\bf REMARK 1.} Of course, regardless  the covariant derivative
$D$ we introduced, in {\em any}
  coordinate patch Euler-Lagrange
equations always read \beq
\label{anterior}\frac{d}{dt}\left(\frac{\partial L}{\partial
\qpun^i}\right) - \frac{\partial L}{\partial q^i}=0\ .\eeq

{\bf REMARK 2.} As well known (e.g. \cite{kn} ), a torsion free
$D$ can always be chosen. It is obvious that for such a
connection, global Euler-Lagrange equations read \beqa
\frac{D}{Dt}\left(\frac{\partial L}{\partial \qpun}\right) -
\frac{DL}{Dq} = 0 . \eeqa So in this case,  the global expression
can be obtained merely  by replacing the usual derivatives by  $D$
in \eq{anterior}.

{\bf REMARK 3.} A similar result holds  for the horizontal
Lagrange-Poincar\'e equations considered in \cite{cmr1,cmr2}. One
of the goals of these references is to analyze the intrinsic
meaning of motion equations for constrained systems with
symmetries. Let us recall that, in such a system, the lagrangian
$L$ and the constrains  remain invariant under the lifting to $TQ$
of a suitable action of a Lie group $G$ on $Q$. A connection $A$,
related to the constrains, is introduced on the principal bundle
$Q\overset{\pi}{\rightarrow} Q/G$. If $\tilde{\mathfrak{g}}$ is
the adjoint bundle to the principal bundle $Q$, the connection $A$
yields an isomorphism $\alpha$ between $TQ/G$ and the Whitney sum
$T(Q/G)\oplus \tilde{\mathfrak{g}}$ in the following way \beq
\alpha_A[q,\qpun]_G= (x,\dot x ,\tilde v)= \pi_*(q,\qpun)\oplus
[q,A(q,\qpun)]_G\ .\eeq Now, one can define  the reduced
Lagrangian $\ell:T(Q/G)\oplus \tilde{\mathfrak{g}}
\rightarrow\mathbb{R} $ as \beq \ell(x,\dot x ,\tilde
v)=L(q,\qpun)\ .\eeq A variation $\delta q$ of a curve in $Q$ is
said to be {\em horizontal} if $A(\delta q)=0$. In this case,  the
corresponding variation $\alpha(\delta q(t))$ of the curve
$\alpha(q(t))$ in $T(Q/G)\oplus \tilde{\mathfrak{g}}$ is
\cite{cmr1} \beq \alpha(\delta q(t))=\delta x \oplus \tilde B
(\delta x,\dot x)\ ,\eeq where $\tilde B$ is the
$\tilde{\mathfrak{g}}$-valued two-form on $Q/G$ defined by \beq
\tilde B([q]_G)(X,Y)=[q,B(X^h(q),Y^h(q))]_G\ ,\eeq with $X^h,Y^h$
the horizontal lifts to $Q$ of $X$ and $Y$, and $B$ the curvature
of the connection $A$.

 The horizontal Lagrange-Poincar\'e equations
for $L$ are defined as the Euler-Lagrange ones for $\ell$
restricted to horizontal variations $\delta q$. A coordinate free
version of them can be written down by introducing an arbitrary
covariant derivative $D$ on $T(Q/G)$ and using the covariant
derivative $\tilde{D}$ induced by $A$ on $\tilde{\mathfrak{g}}$.

Under the implicit assumption that the torsion of $D$ vanishes, it
is shown in \cite{cmr1,cmr2}  that, for horizontal variations
$\delta q$, \beq \delta \int_{t_0}^{t_1}L(q(t),\qpun(t))dt =0\eeq
if and only if the following horizontal Lagrange-Poincar\'e
equations hold \beqa \frac{D}{Dt}\left(\frac{\partial
\ell}{\partial \dot x}\right)(x,\dot x, \tilde v) -
\frac{D\ell}{Dx}(x,\dot x, \tilde v)= -<\frac{\partial
\ell}{\partial \tilde v},\tilde B (x)(\dot x,.) > \ .\eeqa

 It is easy to see that, for an arbitrary
covariant derivative $D$ on $T(Q/G)$, its torsion tensor $T$ must
be taken into account in the previous  equations. Arguing as above
one gets \beqa \frac{D}{Dt}\left(\frac{\partial \ell}{\partial
\dot x}\right)(x,\dot x, \tilde v) - \frac{D\ell}{Dx}(x,\dot x,
\tilde v)= -<\frac{\partial \ell}{\partial \tilde v},\tilde B
(x)(\dot x,.) > +\  T(\qpun,.)\ .\nn \eeqa

Assuming as in \cite{cmr1,cmr2}  the torsion free requirement for
$D$, the last term clearly vanishes and we recover the horizontal
Lagrange-Poincar\'e equations found in those references.

Again, in {\em any} coordinate patch, the expression of horizontal
Lagrange-Poincar\'e equations is independent of the choice of the
covariant derivative $D$.

{\bf REMARK 4.} When considered as a map of the second order
tangent bundle to the cotangent bundle, the Euler-Lagrange
operator  turns out to be  intrinsic without any choice of
connection ( e.g., \cite{mr}). The need for connections appears if
one wants to stay in the framework of tangent bundles, as it is
usually done, and not to deal with second order ones.

\ack This work was partially supported by CONICET, Argentina.

\end{document}